\documentclass[aps,prd,groupedaddress,showpacs]{revtex4}
\usepackage{graphicx}
\usepackage{dcolumn}
\usepackage{bm}
\usepackage{amsmath}
\usepackage{epstopdf}
\usepackage{amsfonts}
\usepackage{amssymb}%

\usepackage[colorlinks=true,citecolor=blue,urlcolor=magenta,breaklinks]{hyperref}

\usepackage{natbib}

\providecommand{\U}[1]{\protect\rule{.1in}{.1in}}

\newcommand{\be}{\begin{equation}}
\newcommand{\en}{\end{equation}}
\newcommand{\bea}{\begin{eqnarray}}
\newcommand{\ena}{\end{eqnarray}}

\begin{document}

\title{Scalar field descriptions of two dark energy models}

\author{Grigorios Panotopoulos}
\email{grigorios.panotopoulos@tecnico.ulisboa.pt}
\affiliation{CENTRA, Instituto Superior T{\'e}cnico,\\ Universidade de Lisboa,
Av. Rovisco Pa{\'i}s 1, Lisboa, Portugal}

\date{\today}

\begin{abstract}
We give a scalar field description of two dark energy parameterizations, and we analyze in detail its cosmology both at the level of background
evolution and at the level of linear perturbations. In particular, we compute the statefinder parameters
and the growth index as functions of the red-shift for both dark energy parameterizations, and the comparison with the $\Lambda CDM$ model as well as
with a few well-known geometrical dark energy models is shown. In addition, the combination parameter $A=f \sigma_8$ of both models is compared against current data.
\end{abstract}


\maketitle

\section{Introduction}

In the end of the 90's the most dramatic discoveries in Particle Physics and Cosmology were from the one hand the neutrino oscillations and on the other hand the current acceleration of the Universe \cite{SN1,SN2}. Nowadays many well established observational data from Astrophysics and Cosmology show that we live in a spatially flat Universe that expands at an accelerating rate \cite{turner}. Dark energy, the fluid component that dominates the evolution of the Universe and drives the current cosmic acceleration, is one of the biggest challenges of modern cosmology, as its nature and origin still remains a mystery. The $\Lambda CDM$ model with a constant equation of state $w=-1$ is the most economical one in excellent agreement with current data. However, given the cosmological constant problems other alternatives with an evolving equation of state have been studied in the literature over the years.

In phenomenological descriptions dark energy is viewed as a perfect fluid with a time varying equation of state $w(a)$, where $a$ is the scale factor. In \cite{Nesseris:2004wj} the authors compared several dark energy parameterizations to supernovae data, and concluded that models that cross the $w=-1$ line have a
better fit to data. Furthermore, in a more recent work it was showed that the cosmic acceleration may have slowed down recently \cite{Magana:2014voa}.
More generally speaking, all dark energy models fall in two broad classes, namely either dynamical dark energy models in which a new dynamical field is introduced to accelerate the Universe \cite{Copeland:2006wr}, or geometrical dark energy models in which a new gravitational theory \cite{Sotiriou:2008rp,DeFelice:2010aj} is assumed that modifies Einstein's general Relativity at large scales.

Many models predict the same expansion history and thus cannot be ruled out by current data. That is why it is important to introduce and study appropriate quantities that can discriminate between several dark energy models. One option is to study how matter perturbations evolve, as its evolution is sensitive to the sound speed of dark energy that differs from one model to another. In particular, a quantity that has been studied a lot over the years is the so called growth index $\gamma$, introduced in \cite{Wang:1998gt} and to be defined later on, and for the $\Lambda CDM$ model has been found to be $\gamma_{\Lambda CDM} = 6/11 \simeq 0.55$ \cite{Nesseris:2007pa,domenico,latino}. Another option is to analyze the so called statefinder parameters $r, s$ expressed in terms of the third derivative of the scale factor $a$ with respect to the cosmic time $t$ as follows \cite{Sahni:2002fz,Alam:2003sc}
\begin{eqnarray}
r & = & \frac{\dddot{a}}{a H^3} \\
s & = & \frac{r-1}{3 (q-\frac{1}{2})}
\end{eqnarray}
where the dot denotes differentiation with respect to the cosmic time, $H=\dot{a}/a$ is the Hubble parameter, and $q=- \ddot{a}/(a H^2)$ is the decelerating parameter. For $\Lambda CDM$ model the statefinder parameters are just constants, $r=1,s=0$. These parameters can be computed within a certain model, their values can be extracted from future observations \cite{SNAP1,SNAP2}, and as we will
show later on can be very different from one model to another even if the two expansion histories are very similar to one another. The statefinder diagnostics has been applied to several dark energy models \cite{diagnostics1,diagnostics2,diagnostics3,diagnostics4,diagnostics5}.

Although it is convenient to study dark energy parameterizations, since for a given $w(a)$ the expansion history of the Universe is known, a more fundamental description is often needed, based on a canonical scalar field for example. However, since in most parameterizations, such as the linear ansatz \cite{linear} or the Linder model \cite{linder}, the equation of state lies outside the interval $[-1,1]$ a scalar field based description is not possible. In this work we wish to study the cosmology of two concrete dark energy parameterizations introduced in \cite{Feng:2012gf} a few years ago, for two reasons. First, as they are relatively new they have not extensively studied yet, and also since the equation of state parameter remains always in the $(-1,1)$ range it can be viewed as a minimally coupled scalar field.

Therefore, the goal of the present article is twofold. First we shall give a scalar field description of the dark energy parameterizations of \cite{Feng:2012gf}, and
then we will analyze its cosmology along the lines discussed previously.
Our work is organized as follows: after this introduction, we present the theoretical framework in section two, and we present our numerical results in the third section. Finally we conclude in section four. We work in natural units in which the speed
of light in vacuum $c$ and the reduced Planck mass $\hbar$ are set equal to unity.

\section{Theoretical framework}

Here we introduce all the necessary ingredients, first for the background evolution and then for the evolution of cosmological
perturbations.

\subsection{Background evolution}

The starting point is four-dimensional Einstein's General Relativity in which gravity is coupled to 3 fluids, namely non-relativistic matter (m),
radiation (r) and dark energy (X). Each fluid is characterized by its energy density $\rho_A$, pressure $p_A=w_A \rho_A$, with $w_A$ being the equation of state parameter of the fluid A, and assuming no interaction between them the continuity equation for each fluid reads
\begin{equation}
\dot{\rho_A}+3 H \rho_A (1+w_A) = 0
\end{equation}
where the dot denotes differentiation with respect to the cosmic time $t$, $H=\dot{a}/a$ is the Hubble parameter, and the index A takes 3 values, $A=m,r,X$.
The equation of state parameter for matter and radiation are $w_m=0$ and $w_r=1/3$ respectively, while for dark energy we shall assume some
parameterization where its equation of state will be a certain function of the red-shift $1+z=a_0/a$, with $a_0$ being today's value of the scale factor $a$.
Furthermore, the evolution of the scale factor is determined by the two Friedmann equations of a flat Robertson-Walker metric
\begin{eqnarray}
H^2 & = & \frac{8 \pi G \rho}{3} \\
\dot{H} & = & - 4 \pi G (\rho + p)
\end{eqnarray}
where $G$ is Newton's constant, $\rho=\rho_m+\rho_r+\rho_X$ is the total energy density, $p=p_m+p_r+p_X$ is the total pressure, and we can
define the total equation of state parameter $w=p/\rho$. It is also convenient to define the normalized energy density
for each fluid, $\Omega_A=\rho_A/\rho$ to be used later on, and the condition $\sum_A \Omega_A = 1$ has to be satisfied.
Using the definitions combined with
the cosmological equations, the deceleration parameter $q$ as well
as the first statefinder parameters $r$ as functions of the red-shift are found to be
\begin{eqnarray}
q(z) & = & -1+(1+z) \frac{E'(z)}{E(z)} \\
r(z) & = & q(z) (2q(z)+1) + (1+z) q'(z)
\end{eqnarray}
while the second statefinder parameter $s$ can be computed accordingly using its definition. The prime denotes differentiation with respect to red-shift,
and $E(z)=H(z)/H_0$ is the dimensionless Hubble parameter versus red-shift,
with $H_0=100 \: h (km \: sec^{-1})/Mpc$ being the Hubble constant.
We have checked that expressions above in the case of the $\Lambda CDM$ model give $r=1,s=0$ as they should. For matter and radiation the normalized energy densities are given by
\begin{eqnarray}
\Omega_m(z) & = & \frac{\Omega_{m,0} (1+z)^3}{E(z)^2} \\
\Omega_r(z) & = & \frac{\Omega_{r,0} (1+z)^4}{E(z)^2}
\end{eqnarray}
where $\Omega_{m,0}, \Omega_{r,0}$ are today's values of the normalized densities for matter and radiation respectively.
Finally, for a given dark energy parameterization $w(z)$, the dimensionless Hubble parameter is given by
\begin{equation}
E(z) = \sqrt{\Omega_{m,0} (1+z)^3+\Omega_{r,0} (1+z)^4+(1-\Omega_{m,0}-\Omega_{r,0}) F(z)}
\end{equation}
where $\Omega_{r,0}=9 \times 10^{-5}$ \cite{Magana:2014voa}, and
where the function $F(z)$ is computed once the dark energy equation of state is given \cite{Nesseris:2004wj}
\begin{equation}
F(z) = exp\left(3 \int_0^z \: dx \: \frac{1+w(x)}{1+x} \right)
\end{equation}
so that $F(0)=1$ and $E(z)=1$ as it should.

It is always desirable to have a more fundamental description, ideally a Lagrangian formulation of the model. If the dark energy equation of state remains always in the range $-1 < w(z) < 1$, the assumed dark energy parameterization can be realized using a canonical scalar field
$\phi$ with an appropriate positive self-interaction potential $V(\phi)$. It is known that a minimally coupled scalar field behaves like a perfect fluid with
pressure $p_{\phi}=(1/2) \dot{\phi}^2-V$ and energy density $\rho_{\phi}=(1/2) \dot{\phi}^2+V$ \cite{Mukhanov:2005sc}. Therefore using the Friedmann equations one can determine both the potential $V(z)$ and the scalar field $\phi(z)$ as functions of red-shift as follows
\begin{equation}
\frac{V(z)}{H_0^2 m_{pl}^2} = \frac{E(z)^2}{8 \pi} (2-q(z)-\Omega_r(z)-\frac{3}{2} \Omega_m(z))
\end{equation}
\begin{equation}
\frac{d(\phi/m_{pl})}{dz} = -\frac{1}{1+z} \sqrt{\frac{1+q(z)-2 \Omega_r(z)-\frac{3}{2} \Omega_m(z)}{4 \pi}}
\end{equation}
with $m_{pl}=1.22 \times 10^{19} GeV$ being the Planck mass. We have checked that the previous expressions in the case of the $\Lambda CDM$ model give $V(z)=constant$ and $\phi'(z)=0$. These equations give the scalar potential $V(\phi)$ in parametric form $\phi(z),V(z)$.

\subsection{Cosmological perturbations}

Now we move on to discuss linear cosmological perturbation theory \cite{Mukhanov:2005sc,dePutter:2010vy,Albarran:2016mdu}. Regarding scalar perturbations, and assuming vanishing anisotropic stress tensor for the fluid components, the metric is characterized by the Bardeen potential $\Psi(\eta, \vec{x})$
\begin{equation}
ds^2 = a(\eta)^2 [-(1+2 \Psi) d\eta^2 + (1-2 \Psi) \delta_{i j} dx^i dx^j]
\end{equation}
with $d\eta=dt/a$ being the conformal time. Furthermore each fluid component is characterized by energy density perturbation $\delta \rho_A (\eta, \vec{x})$, pressure perturbation $\delta p_A (\eta, \vec{x})$ and peculiar velocity potential $v_A (\eta, \vec{x})$ defined from the $0-i$ component of the perturbed stress tensor $\delta T_{A,0}^i=-(\rho_A+p_A) \partial^i v_A$ \cite{Albarran:2016mdu} where $p_A, \rho_A$ are the background quantities. In addition, each fluid component is characterized by two sound speeds that in general are different, namely the adiabatic speed of sound $c_{A,a}^2$, defined by \cite{dePutter:2010vy,Albarran:2016mdu}
\begin{equation}
c_{A,a}^2 = \frac{\dot{p_A}}{\dot{\rho_A}}=w_A-\frac{\dot{w_A}}{3 H (1+w_A)}
\end{equation}
as well as the effective speed of sound in the rest frame of the fluid $c_{A,s}^2$, defined by \cite{dePutter:2010vy,Albarran:2016mdu}
\begin{equation}
\delta p_A = c_{A,s}^2 \delta \rho_A - 3 \mathcal{H} (1+w_A) \rho_A v_A (c_{A,s}^2-c_{A,a}^2)
\end{equation}
where $\mathcal{H} a = da/d\eta$ is the conformal Hubble parameter, and the prime now denotes differentiation with respect to the conformal time. For matter and radiation the two speed of sounds are equal to the equation of state parameter, $c_{m,a}^2=0=c_{m,s}^2$ for matter and $c_{r,a}^2=1/3=c_{r,s}^2$ for radiation. For dark energy one first has to specify the model. If dark energy equation of state $w(z)$ is realized by a canonical scalar field $\phi$ with an appropriate self-interaction potential $V(\phi)$, the effective speed of sound is unity, $c_{X,s}^2=1$ \cite{Erickson:2001bq}.

The perturbed Einstein's equations $\delta G_{\mu \nu}=8 \pi G \delta T_{\mu \nu}$ take the form (in Fourier space) \cite{Albarran:2016mdu}
\begin{eqnarray}
3 \mathcal{H} (\mathcal{H} \Psi+\Psi')+k^2 \Psi & = & -4 \pi G a^2 \delta \rho \\
\mathcal{H} \Psi+\Psi' & = & -4 \pi G a^2 (\rho+p) v \\
\Psi'' + 3 \mathcal{H} \Psi'+\Psi (\mathcal{H}^2+2 \mathcal{H}') & = & 4 \pi G a^2 \delta p
\end{eqnarray}
where $\delta p=\sum_A \delta p_A$ is the total pressure perturbation, $\delta \rho=\sum_A \delta \rho_A$ is the total energy density perturbation, and
$(1+w) v = \sum_A (1+w_A) \Omega_A v_A$ is the total peculiar velocity potential \cite{Albarran:2016mdu}. These are all differential equations for the metric perturbation $\Psi$. However, the stress tensor conservation for each fluid provides us with additional differential equations for the fluid quantities $v$ and density contrast $\delta_A = \delta \rho_A/\rho_A$. Defining the total density contrast $\delta = \delta \rho/\rho=\sum_A \Omega_A \delta_A$, and using as independent variable
$x=-log(1+z)$, so that for any variable $A' = \mathcal{H} A_x$, we finally obtain the following equations for the metric perturbation \cite{Albarran:2016mdu}
\begin{eqnarray}
\Psi_x + \Psi \left(1+\frac{k^2}{3 \mathcal{H}^2} \right) & = & -\frac{\delta}{2} \\
\Psi_x + \Psi & = &  -\frac{3}{2} \mathcal{H} v (1+w)
\end{eqnarray}
as well as the following equations for each fluid component \cite{Albarran:2016mdu}
\begin{eqnarray}
(\delta_r)_x  & = & \frac{4}{3} \left(3 \Psi_x + v_r \frac{k^2}{\mathcal{H}} \right)   \\
(v_r)_x & = & -\frac{1}{\mathcal{H}} \left(\Psi + \frac{\delta_r}{4} \right)  \\
(\delta_m)_x  & = & 3 \Psi_x + v_m \frac{k^2}{\mathcal{H}} \\
(v_m)_x & = & - \left(v_m + \frac{\Psi}{\mathcal{H}} \right) \\
(\delta_X)_x & = & 3 (w_X-c_{X,s}^2) \delta_X + (1+w_X) \left[3 \Psi_x + v_X \left(\frac{k^2}{\mathcal{H}} + 9 \mathcal{H} (c_{X,s}^2-c_{X,a}^2) \right) \right] \\
(v_X)_x & = & v_X (3 c_{X,s}^2-1)-\frac{1}{\mathcal{H}} \left(\Psi +\delta_X \frac{c_{X,s}^2}{1+w_X}  \right)
\end{eqnarray}

Before we go on a comment is in order here. The perturbed Universe is considerably more complicated than the homogeneous one, and depending on the
problem at hand one may have to confront with several difficulties. Just to mention a few, first note that the structure of the previous equations for the dark energy perturbations indicates that for evolving dark energy equations of state $w(a)$ that cross the "phantom divide line", the $1+w_X$ factor in the denominator vanishes. Furthermore, it has been shown that in interacting dark energy models large instabilities that can spoil the growth of the linear perturbations may be present, see e.g. \cite{coupled1,coupled2,coupled3,coupled4,coupled5} and references therein. In the present work, however, we consider non-coupled and non-phantom dark energy models, and therefore none of the aforementioned problems arise here.

Assuming adiabatic initial conditions
\begin{equation}
\frac{\delta_i}{1+w_i} = \frac{\delta_j}{1+w_j}
\end{equation}
for any two fluids $i,j$, the previous differential equations are supplemented with the following initial conditions for the peculiar velocity potentials \cite{Albarran:2016mdu}
\begin{equation}
v_{A,ini} = \frac{\delta_{ini}}{4 \mathcal{H}_{ini}}
\end{equation}
while for the density contrasts we have
\begin{equation}
\delta_{A,ini} = \frac{3}{4} (1+w_{A,ini}) \: \delta_{ini}
\end{equation}
starting from radiation domination where $z_{ini}=10^6$ or $x_{ini}=-13.81$. For the Hubble constant $H_0$ we have used the Planck 2015 results \cite{planck}. The growth index $\gamma$ is defined through the relation below \cite{Linder1,Linder2,Linder3}, and for more recent discussions see e.g. \cite{Ballesteros:2008qk,growth1,growth2,growth3}
\begin{equation}
\frac{d (ln\delta_m)}{d (lna)} = f = \Omega_m^\gamma
\end{equation}
Therefore the function $f$ is computed first from the matter energy density contrast, and then the growth index is given by
\begin{equation}
\gamma = \frac{ln(f)}{ln(\Omega_m)}
\end{equation}
We note in passing that one can integrate the equations for the perturbations using one of the publicly available computer codes, such as CMBFAST \cite{cmbfast}
or CAMB \cite{camb} or CMB-easy \cite{cmbeasy,doran}. However, since we don't need to compute the temperature anisotropies here, we prefer to numerically integrate the equations using a Mathematica \cite{wolfram} file as was done in \cite{leandros}.

\section{Numerical results}

We can now analyze any dark energy model with a given parameterization $w(z)$. Here we shall consider the following two \cite{Feng:2012gf}
\begin{eqnarray}
wI(z) & = & w_0 + w_1 \frac{z}{1+z^2} \\
wII(z) & = & w_0 + w_1 \frac{z^2}{1+z^2}
\end{eqnarray}
where the two constants $w_0, w_1$ are determined upon comparison with supernovae data. For the first model $w(z=0)=w(z \gg 1)=w_0$, while
for the second model $w(z=0)=w_0$ and $w(z \gg 1)=w_0+w_1$. For these parameterizations, the function $f(z)$ that determines the dimensionless
Hubble parameter is found to be \cite{Magana:2014voa}
\begin{equation}
F(z) = exp\left(\frac{3 w_1}{2} arctan(z) \right) (1+z)^{3 (1+w_0-\frac{w_1}{3})} (1+z^2)^{\frac{3 w_1}{4}}
\end{equation}
for the first model, and
\begin{equation}
F(z) = exp\left(-\frac{3 w_1}{2} arctan(z) \right) (1+z)^{3 (1+w_0+\frac{w_1}{3})} (1+z^2)^{\frac{3 w_1}{4}}
\end{equation}
for the second model. Using supernovae data from the Union 2.1 compilation \cite{union}, the authors of \cite{Magana:2014voa} found the following values
\begin{eqnarray}
w_0 & = & -0.965 \\
w_1 & = & 0.388 \\
\Omega_{m,0} & = & 0.23
\end{eqnarray}
for the first model, and
\begin{eqnarray}
w_0 & = & -0.942 \\
w_1 & = & 0.531 \\
\Omega_{m,0} & = & 0.239
\end{eqnarray}
for the second model.

The equation of state versus red-shift is shown in Fig. 1 for both models. As one can see, it always remains in the range $(-1,1)$. Given the equations presented in Section 2 one can easily compute the potential of the canonical scalar field $V(\phi)$ that gives rise to the same expansion history, and it is shown in Fig. 2.
To produce the plot, an initial condition for the scalar field is required, and we have assumed its present value to be $\phi(z=0)=0.5 m_{pl}$.
Furthermore, one can also compute the expansion history $E(z)$ as well as the deceleration parameter $q(z)$, and it turns out that they are very similar to one another. However, things completely change if we compare the models I and II using the statefinder parameters $r, s$. In Fig. 3 we show the first parameter $r$ versus red-shift for both models, while in Fig. 4 the second parameter $s$ versus red-shift for both models is shown.

Next we consider linear cosmological perturbations and the evolution of the functions $f(z), \gamma(z)$. Fig. 5 and 6 show $\gamma$ and $f$ respectively for model I and 3 different scales $k$ of the linear regime. Similarly, Fig. 7 and 8 show the same quantities for model II and the same 3 values of $k$. For comparison we show in the same plot (dotted curves) the corresponding quantities of the $\Lambda CDM$ model. In fig. 9 we compare the growth index of models I and II for scale $k=0.005 h Mpc^{-1}$ with the Dvali-Gabadadze-Porrati(DGP) brane model \cite{dgp} as well as two modified gravity $f(R)$ models \cite{HS,starobinsky}. Assuming a parameterization for the growth index of the form $\gamma(z)=\gamma_0+(\gamma_1 z)/(1+z)$, the free parameters $\gamma_0, \gamma_1$ for the $f(R)$ models are given in Table III of \cite{basilakos1}, while the theoretical value of the growth index for the DGP model has been found to be $\gamma_{DGP}=11/16 \simeq 0.6875$ \cite{gammadgp}.

Finally in Fig. 10 we compare the prediction of the models to available data regarding the combination parameter $A(z)=\sigma_8(z) f(z)$, where the rms fluctuation $\sigma_8(z)$ is related to the matter energy density contrast by \cite{Nesseris:2007pa,Albarran:2016mdu}
\begin{equation}
\sigma_8(z)= \frac{\delta_m(z)}{\delta_m(0)} \sigma_8(z=0)
\end{equation}
evaluated at the scale $k_{\sigma_8}=0.125 \: h Mpc^{-1}$ \cite{Albarran:2016mdu}, and we use the Planck results $\sigma_8(z=0) \sqrt{\Omega_{m,0}} \simeq 0.46$ \cite{planck}. The data points with the error bars as well as the relevant references can be seen in Table II of \cite{Albarran:2016mdu}.
Fig. 10 looks very similar to analogous figures produced in other related works, such as \cite{Albarran:2016mdu,alcaniz,basilakos2}.
The red curve corresponds to model I, while the blue curve corresponds to model II.
As we can see, at small red-shifts the two curves coincide, while at $z \simeq 0.7$ they become distinguishable with the red curve being slightly above the blue one.

\begin{figure}[ht!]
\centering
\includegraphics[scale=0.90]{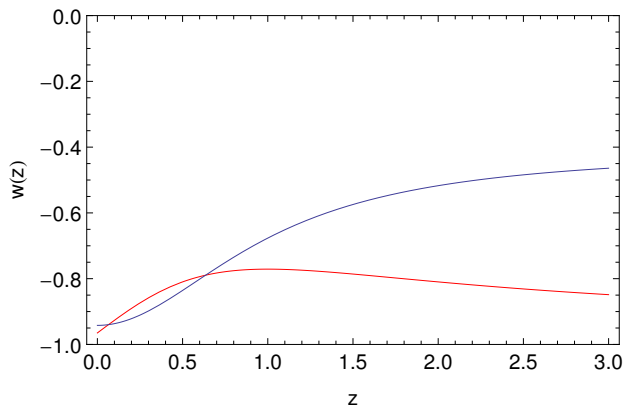}
\caption{Equation of state parameter versus red-shift for models I and II. The red curve corresponds to model I, while the blue curve corresponds to model II.}
\label{fig:1} 	
\end{figure}

\begin{figure}[ht!]
\centering
\includegraphics[scale=0.90]{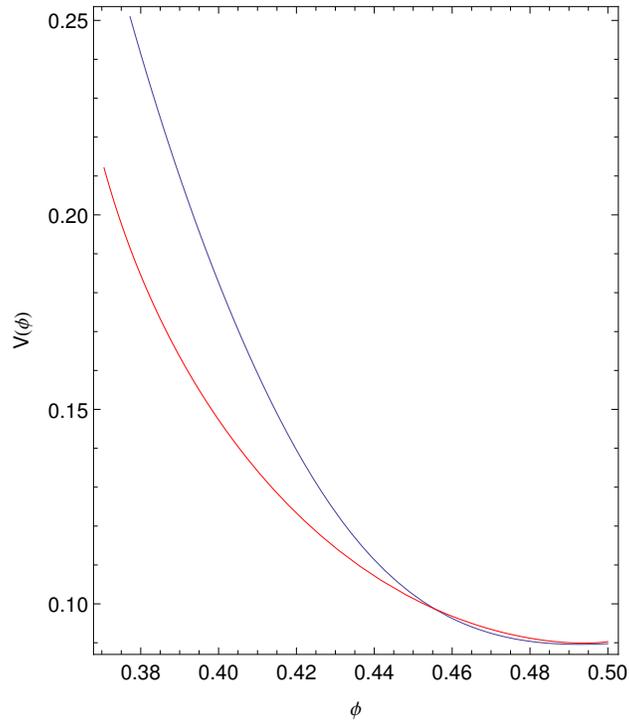}
\caption{Scalar potential (in units of $H_0^2 m_{pl}^2$) as a function of the canonical scalar field (in units of Planck mass) for models I and II. The red curve corresponds to model I, while the blue curve corresponds to model II. We have assumed that the present value of the scalar field is $\phi(z=0)=0.5 m_{pl}$.}
\label{fig:2} 	
\end{figure}

\begin{figure}[ht!]
\centering
\includegraphics[scale=0.90]{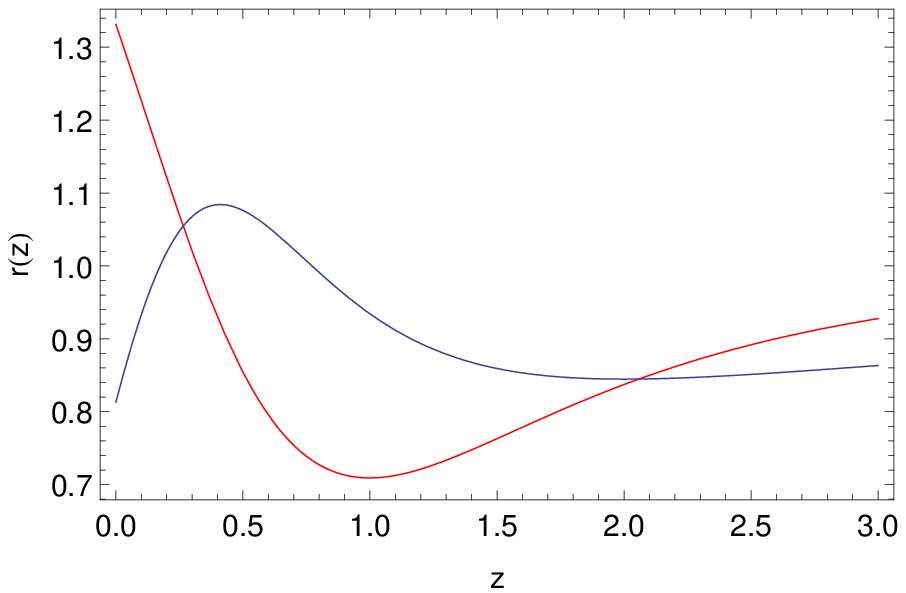}
\caption{The first statefinder parameter $r$ versus red-shift for models I and II. The red curve corresponds to model I, while the blue curve corresponds to model II.}
\label{fig:3} 	
\end{figure}

\begin{figure}[ht!]
\centering
\includegraphics[scale=0.90]{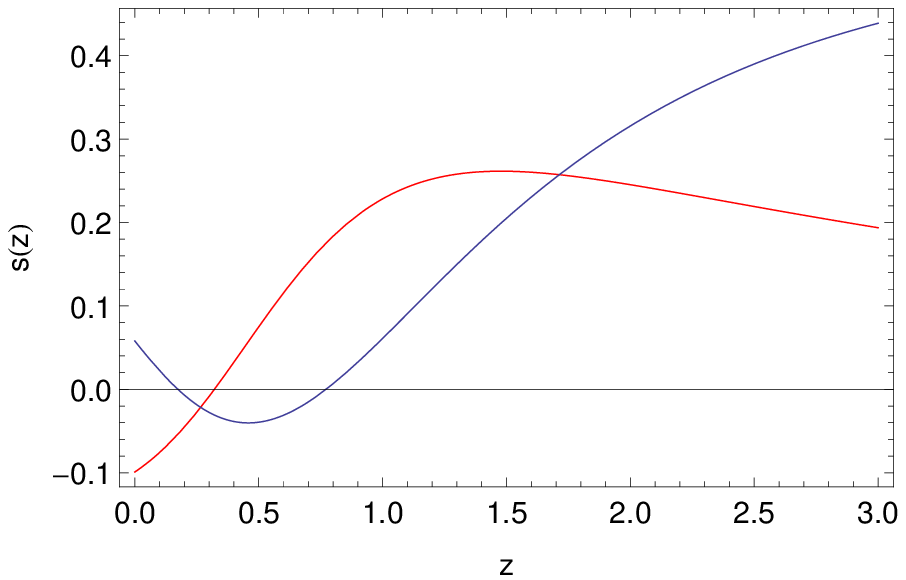}
\caption{The second statefinder parameter $s$ versus red-shift for models I and II. The red curve corresponds to model I, while the blue curve corresponds to model II.}
\label{fig:4} 	
\end{figure}

\begin{figure}[ht!]
\centering
\includegraphics[scale=0.90]{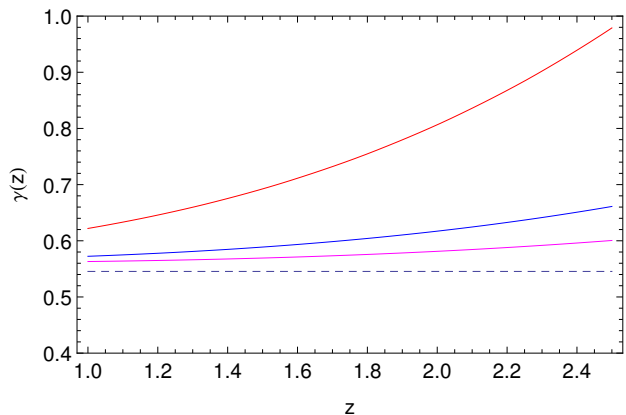}
\caption{Growth index $\gamma$ versus red-shift for model I and for 3 different scales, $k=2 \times 10^{-3} h Mpc^{-1}$ (red), $k=4 \times 10^{-3} h Mpc^{-1}$ (blue), $k=6 \times 10^{-3} h Mpc^{-1}$ (magenta). The dotted line corresponds to the $\Lambda CDM$ model.}
\label{fig:5} 	
\end{figure}

\begin{figure}[ht!]
\centering
\includegraphics[scale=0.90]{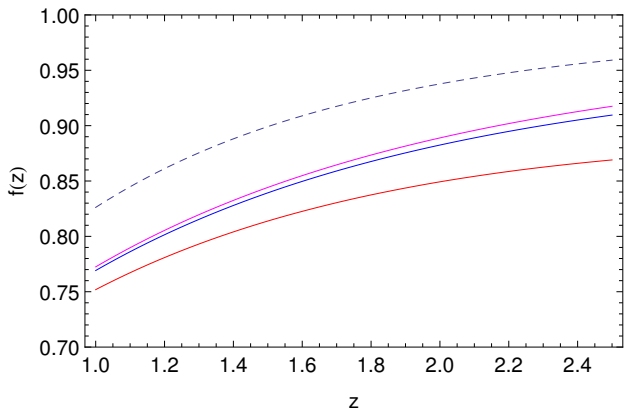}
\caption{The function f versus red-shift for model I and for the 3 different scales considered in Fig. 5. The dotted curve corresponds to the $\Lambda CDM$ model.}
\label{fig:6} 	
\end{figure}

\begin{figure}[ht!]
\centering
\includegraphics[scale=0.90]{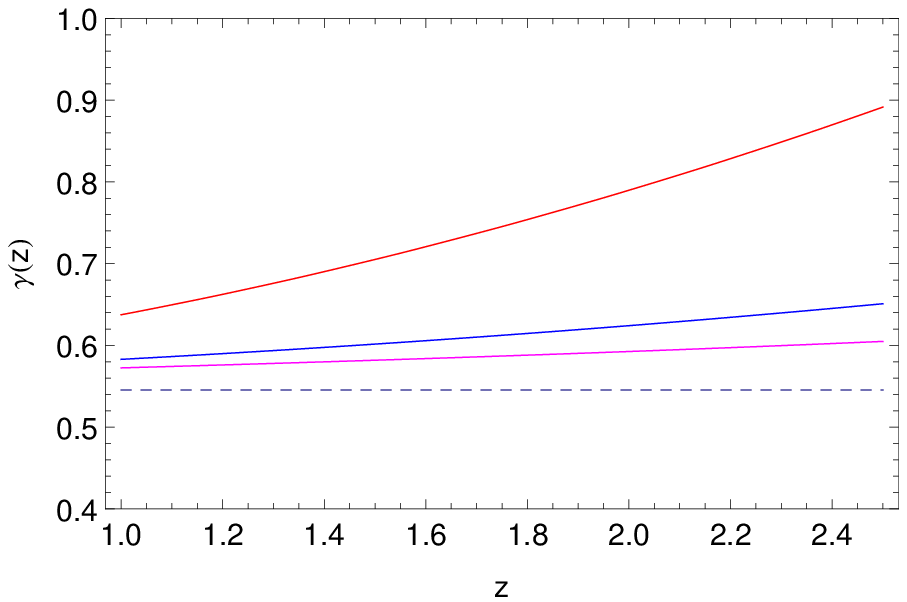}
\caption{Same as Fig. 5 but for model II}
\label{fig:7} 	
\end{figure}

\begin{figure}[ht!]
\centering
\includegraphics[scale=0.90]{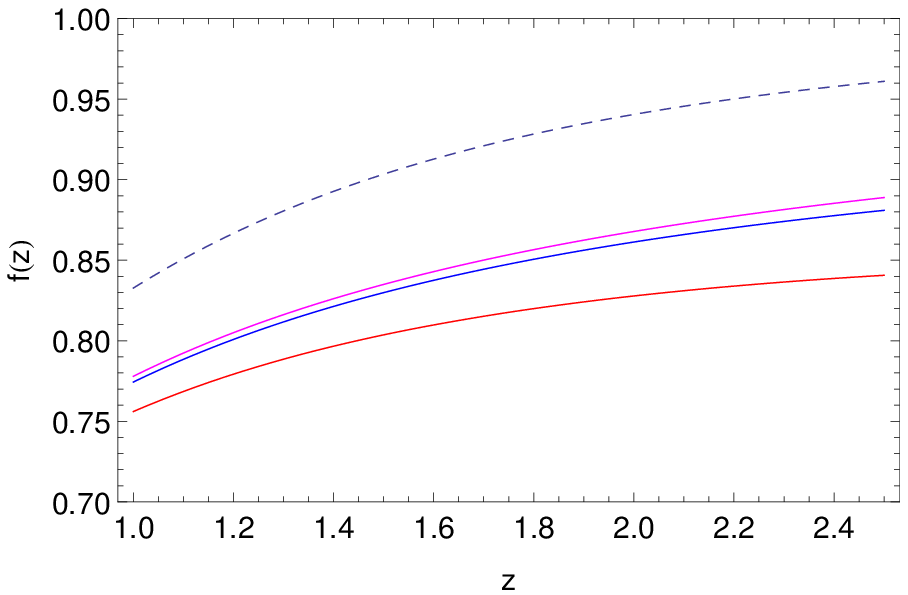}
\caption{Same as fig. 6 but for model II}
\label{fig:8} 	
\end{figure}

\begin{figure}[ht!]
\centering
\includegraphics[scale=0.90]{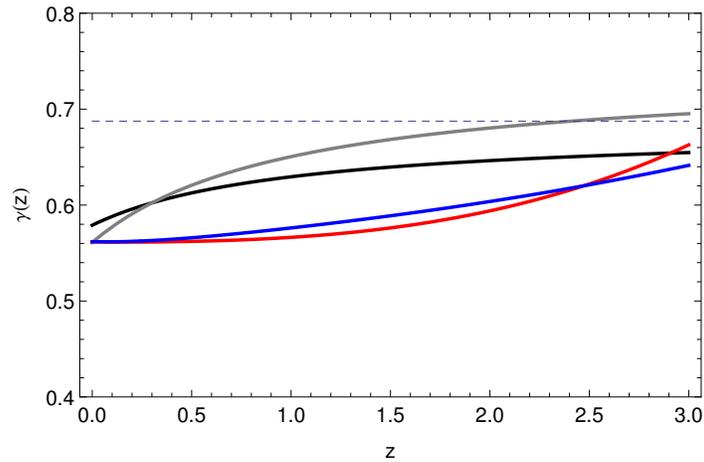}
\caption{Growth index $\gamma$ versus red-shift for models I (red) and II (blue) for scale $k=0.005 h Mpc^{-1}$, for the DGP model \cite{dgp} (dashed), and two modified gravity models (black for \cite{HS} and gray for \cite{starobinsky}).}
\label{fig:9} 	
\end{figure}

\begin{figure}[ht!]
\centering
\includegraphics[scale=0.90]{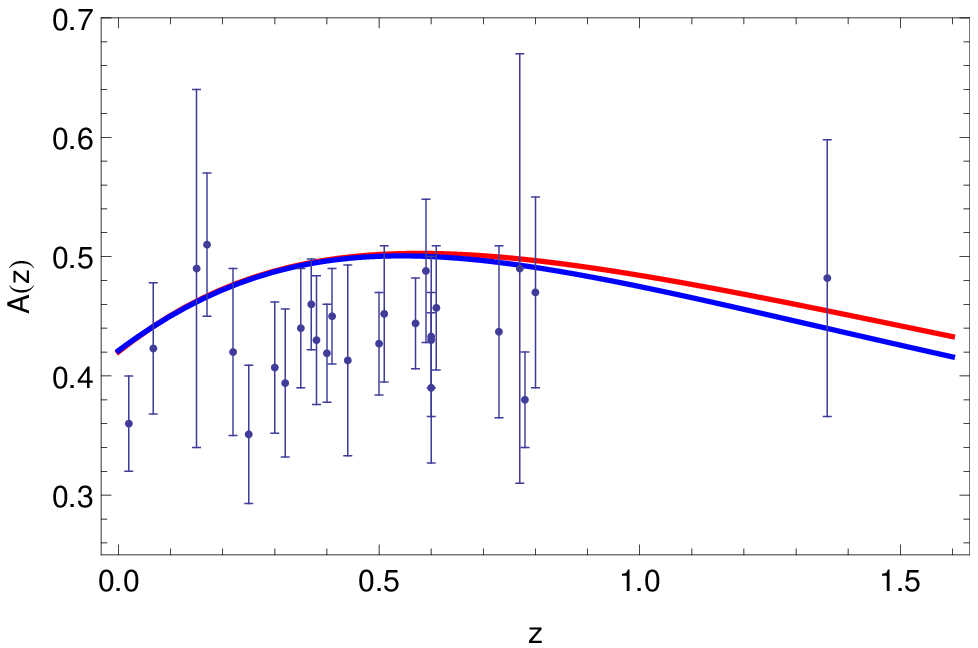}
\caption{Comparison between observational data (the error bars are shown too) and the prediction of the models I and II for $A(z)=\sigma_8(z) f(z)$ versus red-shift. The red curve corresponds to model I while the blue curve corresponds to model II.}
\label{fig:10} 	
\end{figure}

\section{Conclusions}

To summarize, in the present article we have analyzed in detail the cosmology of two dark energy parameterizations both at the level of background evolution and at the level of linear perturbations. The dark energy equation of state is assumed to be a certain function of the red-shift, and it is characterized by two parameters that previously were determined upon comparison with the Union 2.1 compilation supernovae data. The dark energy parameterizations studied here can be realized by a minimally coupled scalar field, and we have computed its appropriate self-interaction potential. At the level of background evolution we have computed the statefinder parameters $r, s$ versus red-shift, while at the level of linear cosmological perturbation theory we have computed $f=(a/\delta_m) d \delta_m/da$ as well as the growth index $\gamma=ln(f)/ln(\Omega_m)$ as functions of the red-shift for both dark energy parameterizations, and for 3 different scales $k$. We have produced several plots in which our main numerical results are shown, and the comparison with the $\Lambda CDM$ model as well as with a few well-known geometrical dark energy models is shown. 
We have also computed the combination parameter $A=f \sigma_8$ for both models, and we have compared the theoretical predictions against available observational data.


\begin{acknowledgments}
The author wishes to thank the anonymous reviewer for valuable comments and suggestions, and is grateful
to M.~Bouhmadi-L{\'o}pez for enlightening discussions.
Work supported from "Funda{\c c}{\~a}o para a Ci{\^e}ncia e Tecnologia".
\end{acknowledgments}


\end{document}